\documentclass[11pt,conference]{IEEEtran}
\usepackage{times}
\usepackage{epsfig}
\usepackage{amsmath}
\usepackage{amsfonts}
\usepackage{amssymb}
\usepackage{amstext}
\usepackage{latexsym}
\usepackage{color}
\usepackage{ifthen}
\usepackage{multirow}
\usepackage{subfigure}
\newcommand{\be}{\begin{equation}}
\newcommand{\ee}{\end{equation}}
\newcommand{\bear}{\begin{eqnarray}}
\newcommand{\eear}{\end{eqnarray}}
\newcommand{\bears}{\begin{eqnarray*}}
\newcommand{\eears}{\end{eqnarray*}}
\newcommand{\bi}{\begin{itemize}}
\newcommand{\ei}{\end{itemize}}
\newcommand{\ben}{\begin{enumerate}}
\newcommand{\een}{\end{enumerate}}

\newcommand{\beq}{\begin{equation}}
\newcommand{\eeq}{\end{equation}}

\newcommand{\lp}{ \left(}
\newcommand{\rp}{ \right)}

\newtheorem{theorem}{Theorem}[section]

\newtheorem{defn}[theorem]{Definition} 
\newtheorem{lemma}[theorem]{Lemma}

\newcommand{\Gbf}{\mbox{${\bf G }$} }

\newcommand{\Sbf}{\mbox{${\bf S }$} }

\newcommand{\FF}{\mbox{$\mathbb{F}$} }

\newcommand{\SNR}{{\sf SNR}}

\begin{document}

\title{A Deterministic Approach to \\ Wireless Relay Networks}

\author{\authorblockN{Amir Salman Avestimehr}
\authorblockA{Wireless Foundations\\
 UC Berkeley,\\
 Berkeley, California, USA.\\
Email: {\sffamily avestime@eecs.berkeley.edu}} \and
\authorblockN{Suhas
N. Diggavi}
\authorblockA{School of Computer and\\
Communication Sciences, EPFL,\\
 Lausanne, Switzerland.\\
Email: {\sffamily suhas.diggavi@epfl.ch}}
\and
\authorblockN{David N C. Tse}
\authorblockA{ Wireless Foundations\\
 UC Berkeley,\\
 Berkeley, California, USA.\\
Email: {\sffamily dtse@eecs.berkeley.edu}}
 }

% make the title area
\maketitle

\begin{abstract}
We present a deterministic channel model which captures several key
features of multiuser wireless communication. We consider a model
for a wireless network with nodes connected by such deterministic
channels , and present an exact characterization of the end-to-end capacity when there is a
single source and a single destination and an arbitrary number of
relay nodes. This result is a natural generalization of the max-flow min-cut theorem for wireline networks. Finally to demonstrate the connections between deterministic model and Gaussian model, we look at two examples: the single-relay channel and the diamond network. We show that in each of these two examples, the capacity-achieving scheme in the corresponding deterministic model naturally suggests a scheme in the Gaussian model that is within 1 bit and 2 bit respectively from cut-set upper bound, for all values of the channel gains. This is the first part of a two-part paper; the sequel \cite{allerton_paper_part_two} will focus on the proof of the max-flow min-cut theorem of a class of deterministic networks of which our model is a special case.

\end{abstract}

%First part of this paper includes:
%1-Introduction
%2-Deterministic Wireless Network Model
%3-Single-Source, Single Destination Network and its Capacity

%\input{paper_one_first_part}

\section{Introduction}
\label{sec:intro}

Two fundamental features distinguish wireless communication from
 wireline communication:
\begin{itemize}

\item first, the {\em broadcast} nature of  wireless communication;
  wireless users communicate over the
air and signals from any one transmitter is heard by multiple nodes
with possibly different signal strengths.

\item second, the {\em superposition} nature; a wireless node
receives signals from multiple simultaneously transmitting nodes,
with the received signals all superimposed on top of each other.
\end{itemize}

Because of these two effects, links in a wireless network are never
isolated but instead interact in seemingly complex
 ways. This is quite unlike  the wired world
where each transmitter-receiver pair can often be thought of as
isolated point-to-point links.

The multiuser Gaussian channel model is the standard one used in
information theory to capture these two effects: signals get
attenuated by complex gains and added together with Gaussian noise
at each receiver (the Gaussian noises at different receivers being
independent of each other.). Unfortunately, except for the simplest
networks such as the one-to-many Gaussian broadcast channel and the
many-to-one Gaussian multiple access channel, the capacity region of
most Gaussian networks is unknown. For example, even the capacity of
the simplest Gaussian relay network, with a single source, single
destination and single relay, is an open question.

To make further progress, in this paper we present a new multiuser
channel model which is analytically simpler than Gaussian models but
yet still captures the two key features of wireless communication of
broadcast and superposition. The key feature of this model is that
the channels are {\em deterministic}: the signal received at a node
in the network is a (deterministic) function of the transmitted
signals. This model is a good approximation of the corresponding
multiuser Gaussian model under two assumptions that are quite common
in many wireless communication scenarios:

\begin{itemize}

\item the additive noise at each receiver is small compared to the
strength of the signals received from the transmitters (high SNR
regime)

\item the signals from different nodes in the network can be
received at very different power at a given receiver (high dynamic
range of received signals)

\end{itemize}

Essentially, this class of deterministic models allow us to focus on
the interaction between the signals transmitted from the different
nodes of the network rather than the noise. In this paper we first introduce and motivate the deterministic model through three basic examples: point-to-point, broadcast and multiple-access channels. Then we consider a network with a single source and a single destination but with arbitrary number of relay nodes, all connected by such deterministic channels. The cut-set bound on the end-to-end capacity of such networks can actually be achieved, the proof of which can be found in the sequel \cite{allerton_paper_part_two}. Finally to demonstrate the connections between deterministic model and Gaussian model, we look at two examples: The single relay channel and the Diamond network. We show that the capacity-achieving schemes in the corresponding deterministic model naturally suggest schemes whose performance is "close" to the cut-set upper bound in the Gaussian model. More specifically, we show that in the single-relay network, the gap is at most 1 bit/s/Hz, and in the Diamond network, the gap is at most 2 bit/s/Hz. The gaps hold for {\em all} values of the channel gains and are relevant particularly when the SNR is high and the capacity is large.

%
%To show the power of this class of deterministic models, we consider
%a network with a single source and a single destination but with
%arbitrary number of relay nodes, all connected by such deterministic
%channels. We show that the mincut bound on the end-to-end capacity
%of such networks can actually be achieved; moreover, it is the same
%as the max-flow min-cut solution of a certain naturally related {\em
%wireline} network with orthogonal non-interacting links.
%
%While it is possible to make precise in what sense is the capacity
%of such a deterministic network an approximation to the (unknown)
%capacity of the corresponding Gaussian network, this will be done
%only in the full paper \cite{ADT07}. In this conference paper, the
%relationship between the two models will only be discussed
%intuitively. This discussion will serve as partial justification as
%to why the class of deterministic models we consider here is
%interesting.

\section{A Deterministic Model for Wireless Networks}
\label{sec:DetermisticModel}

In this section we introduce a \textit{ deterministic model} for wireless networks. First we motivate this model by looking at the following three examples:
\begin{enumerate}
    \item Point-to-point channel
    \item Broadcast Channel (BC)
    \item Multiple access channel (MAC)
\end{enumerate}
Through each example we will discuss how the proposed deterministic model captures the fundamental aspects of wireless channels.

\subsection{Point-to-Point}
\begin{figure*}%[htp]
     \centering
     %%\subfigure[Compact representation]{
      %% \input{WirelessGraph.pstex_t}

   %%  \hspace{.3in}
    % \subfigure[Deterministic relay network]{
       \input{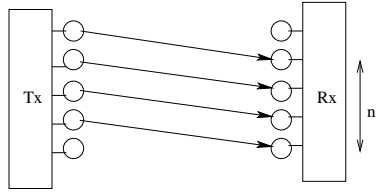}
 \caption{Pictorial representation of the deterministic model for point-to-point channel} \label{fig:p2p_det}
\end{figure*}

Consider an AWGN channel, \beq \label{eq:p2p} y=hx+z \eeq where $z \sim \mathcal{CN} (0,1)$. There is also an average power constraint $E[|x|^2]\leq 1$ at the transmitter. In this paper we normalize both transmit power and noise power to be equal to 1 and capture the signal-to-noise ratio ($\SNR$) in terms of channel gains. So we model $h$ as a \emph{fixed} number representing the channel gain, hence \beq h=\sqrt{\SNR} \eeq  It is well known that the capacity of this point-to-point channel is
\begin{eqnarray} \label{eq:p2p_Gaussian_cap} C_{\text{AWGN}}& =& \log \lp 1+ \SNR \rp  \end{eqnarray}
To get an intuitive understanding of this capacity formula lets write the received signal in equation (\ref{eq:p2p}), $y$, in terms of binary expansions of $x$ and $z$. For simplicity assume $x$ and $z$ are real numbers, then we have
\beq y= 2^{\frac{1}{2}\log \SNR} \sum_{i=1}^{\infty}x(i)2^{-i}  + \sum_{i=-\infty}^{\infty}z(i)2^{-i}\eeq
To simplify the effect of background noise assume it has a peak power equal to 1. Then we can write
 \begin{eqnarray} y&=& 2^{\frac{1}{2}\log \SNR} \sum_{i=1}^{\infty}x(i)2^{-i} + \sum_{i=1}^{\infty}z(i)2^{-i}  \end{eqnarray} or,
\beq y \approx 2^{n} \sum_{i=1}^{n}x(i)2^{-i} + \sum_{i=1}^{\infty}\lp x(i+n)+z(i) \rp 2^{-i} \eeq where $n= \lceil \frac{1}{2} \log \SNR \rceil$. Therefore if we just ignore the 1 bit of the carry-over from the second summation ($\sum_{i=1}^{\infty}\lp x(i+n)+z(i) \rp 2^{-i}$) to the first summation ($2^{n} \sum_{i=1}^{n}x(i)2^{-i}$) we can intuitively model a point-to-point Gaussian channel as a pipe that truncates the transmitted signal and only passes the bits that are above noise level. Therefore think of transmitted signal $x$ as a sequence of bits at different signal levels, with the highest signal level  in $x$ being the most significant bit (MSB) and the lowest level  being the least significant bit (LSB). In this simplified model the receiver can see the $n$ most significant bits of $x$ without any noise and the rest are not seen at all. Clearly there is a correspondence between $n$ and $\SNR$ in dB scale,
\beq n \leftrightarrow \lceil \log  \SNR \rceil \eeq  note that a factor of $\frac{1}{2}$ is needed in the case of AWGN channel with real signals rather than complex signals. As we notice in this simplified model there is no background noise any more and hence we call it a \emph{deterministic model}. Pictorially the deterministic model corresponding to the AWGN channel is shown in figure \ref{fig:p2p_det}. In this figure at the transmitter there are several small circles. Each circle represents a signal level and a binary digit can be put for transmission at each signal level. Depending on $n$, which represents the channel gain in dB scale, the transmitted bits at first $n$ signal levels will be received clearly at the destination. However the bits at other signal levels will not go through the channel. In analogy to the AWGN channel the first $n$ bits are those that are above noise level and the remaining are the ones that are below noise level. Therefore if transmit signal, $\mathbf{x}$, is a binary vector of length $q$, then deterministic channel only delivers its first $n$ elements to the destination. We can algebraically write this input-output relationship by shifting $\mathbf{x}$ down by $q-n$ elements or more precisely
\beq \mathbf{y}={\bf S^{q-n}}\mathbf{x} \eeq
where $\mathbf{x}$ and $\mathbf{y}$ are binary vectors of length $q$ denoting transmit and received signals respectively and $\Sbf$ is the $q \times q$ shift matrix,
\beq \Sbf=\left(
         \begin{array}{ccccc}
           0 & 0 & 0 & \cdots & 0 \\
           1 & 0 & 0 & \cdots & 0 \\
           0 & 1 & 0 & \cdots & 0 \\
           \vdots & \ddots & \ddots & \ddots & \vdots \\
           0 & \cdots & 0 & 1 & 0 \\
         \end{array}
       \right)
 \eeq

Clearly the capacity of this deterministic point-to-point channel is
\beq C_{\text{det}}=n \eeq where $n= \lceil \log  \SNR \rceil$. It is interesting to note that this is a within-one-bit approximation of the capacity of the AWGN channel.

\subsection{Broadcast Channel (BC):}
\begin{figure*}%[htp]
     \centering
     \subfigure[Pictorial representation of the deterministic model for Gaussian BC]{
       \input{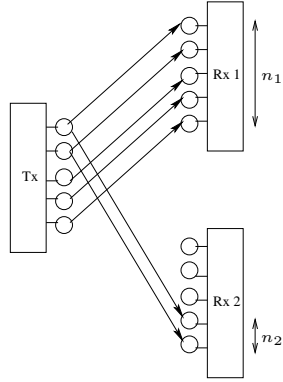} }
    \hspace{.5in}
    \subfigure[Capacity region of Gaussian BC channel (solid line). Capacity region of deterministic BC channel (dashed line)]{
       \input{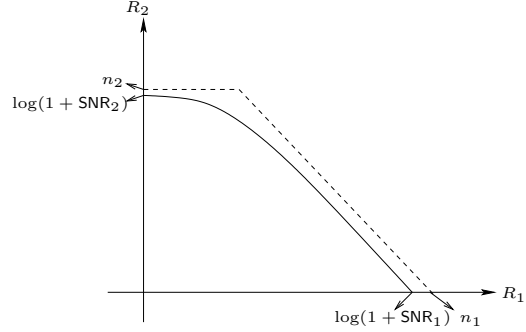}}
 \caption{Pictorial representation of the deterministic model for Gaussian BC is shown in (a). Capacity region of Gaussian and deterministic BC are shown in (b)} \label{fig:bc}
\end{figure*}

Based on the intuition obtained so far, it is straightforward to think of a deterministic model for the Gaussian broadcast channel. Assume there are only two receivers. The received $\SNR$ at receiver $i$ is denoted by $\SNR_i$ for $i=1,2$. Without loss of generality assume $\SNR_2 \le \SNR_1$. Consider the binary expansion of the transmitted signal, $x$. Then we can deterministically model the Gaussian BC channel as the following
\begin{itemize}
  \item Receiver 2 (weak user) receives only the first $n_2$ bits in the binary expansion of $x$. Those bits are the ones that arrive above noise level
  \item Receiver 1 (strong user) receives the first $n_1$ ($n_1>n_2$) bits in the binary expansion of $x$. Clearly these bits contain what receiver 1 gets
\end{itemize}
Pictorially the deterministic model for a Gaussian BC channel is shown in figure \ref{fig:bc} (a). In this particular example $n_1=5$ and $n_2=2$, therefore both users receive the first two most significant bits of the transmitted signal. However user 1 (strong user) receives additional three bits from the next three signal levels of the transmitted signal. There is also the same correspondence between $n$ and channel gains in dB,
\beq n_i \leftrightarrow \lceil \log  \SNR_i \rceil , \quad i=1,2 \eeq
To understand how closely we are modeling the Gaussian BC channel, the capacity region of Gaussian BC channel and deterministic BC channel are shown in figure \ref{fig:bc} (b). In fact it is easy to verify that these regions are within one bit per user of each other (\emph{i.e.} if a pair $(R_1,R_2)$ is  in the capacity region of the deterministic BC then there is a pair within one bit per component of $(R_1,R_2)$ that is in the capacity region of the Gaussian BC).

%\begin{figure*}%[htp]
%     \centering
%     %%\subfigure[Compact representation]{
%      %% \input{WirelessGraph.pstex_t}
%
%   %%  \hspace{.3in}
%    % \subfigure[Deterministic relay network]{
%       \input{bc.pstex_t}
% \caption{Pictorial representation of the deterministic model for Gaussian BC} \label{fig:det_bc}
%\end{figure*}
%
% \begin{figure*}%[htp]
%     \centering
%     %%\subfigure[Compact representation]{
%      %% \input{WirelessGraph.pstex_t}
%
%   %%  \hspace{.3in}
%    % \subfigure[Deterministic relay network]{
%       \input{Gaussian_bc_cap.pstex_t}
% \caption{Capacity region of Gaussian BC channel (solid line). Capacity region of deterministic BC channel (dashed line)} \label{fig:Gaussian_bc_cap}
%\end{figure*}

\subsection{Multiple Access Channel (MAC):}
\begin{figure*}%[htp]
     \centering
     \subfigure[Pictorial representation of the deterministic MAC.]{
       \input{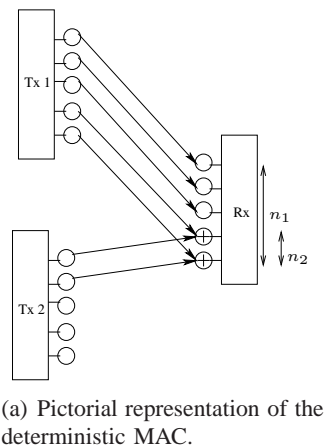}}
    \hspace{.5in}
    \subfigure[Capacity region of Gaussian MAC. (solid line). Capacity region of deterministic MAC.(dashed line)]{
       \input{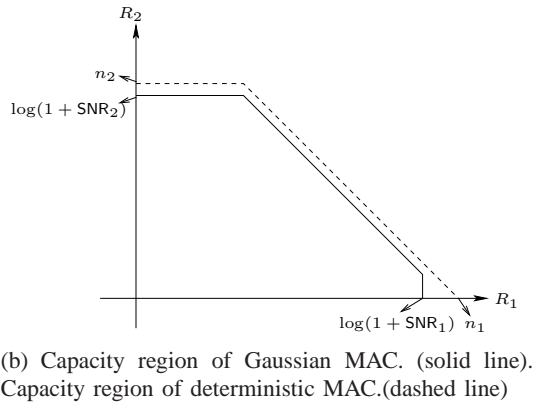}}
 \caption{Pictorial representation of the deterministic MAC is shown in (a). Capacity region of Gaussian and deterministic MACs are shown in (b)} \label{fig:mac}
\end{figure*}

Consider a two user Gaussian MAC channel:
\beq y=h_1 x_1+h_2 x_2+z \eeq
where $z \sim \mathcal{CN} (0,1)$. There is also an average power constraint equal to 1 at both transmitters. The channel gains are
\beq h_i=\sqrt{\SNR_i}, \quad i=1,2 \eeq Without loss of generality assume $\SNR_2 < \SNR_1$.
The capacity region of this channel is well-known to be the set of non-negative pairs $(R_1,R_2)$ satisfying \begin{eqnarray}R_i & \leq & \log (1+\SNR_i), \quad i=1,2 \\
R_1+R_2 & \leq & \log (1+\SNR_1+\SNR_2) \end{eqnarray}  This region is plotted with solid line in figure \ref{fig:mac} (b).

To intuitively understand what happens in a Gaussian MAC channel we write the received signal, $y$, in terms of the binary expansions of $x_1$, $x_2$ and $z$. For simplicity assume $x_1$, $x_2$ and $z$ are all real numbers, then we have
{\footnotesize \beq y= 2^{\frac{1}{2}\log \SNR_1} \sum_{i=1}^{\infty}x_1(i)2^{-i} + 2^{\frac{1}{2}\log \SNR_2} \sum_{i=1}^{\infty}x_2(i)2^{-i} + \sum_{i=-\infty}^{\infty}z(i)2^{-i}\eeq }
To simplify the effect of background noise assume it has a peak power equal to 1. Then we can write
{\footnotesize \beq y= 2^{\frac{1}{2}\log \SNR_1} \sum_{i=1}^{\infty}x_1(i)2^{-i} + 2^{\frac{1}{2}\log \SNR_2} \sum_{i=1}^{\infty}x_2(i)2^{-i} + \sum_{i=1}^{\infty}z(i)2^{-i}\eeq }
or,
{\footnotesize \begin{eqnarray} \nonumber y& \approx & 2^{n_1} \sum_{i=1}^{n_1-n_2}x_1(i)2^{-i} + 2^{n_2} \sum_{i=1}^{n_2}\lp x_1(i+n_1-n_2)+x_2(i)\rp 2^{-i} \\ &&+ \sum_{i=1}^{\infty}\lp x_1(i+n_1)+x_2(i+n_2)+ z(i) \rp 2^{-i}\end{eqnarray} }
where $n_i= \lceil \frac{1}{2} \log  \SNR_i \rceil$ for $i=1,2$. Therefore based on the intuition obtained from the point-to-point AWGN channel, we can approximately model a MAC channel as follows
\begin{itemize}
    \item That part of $x_1$ that is above $\SNR_2$ ($x_1(i)$, $1 \leq i \leq n_1-n_2$) is received clearly without any interference from $x_2$
    \item The remaining part of $x_1$ that is above noise level ($x_1(i)$, $n_1-n_2 < i \leq n_1$) and that part of $x_2$ that is above noise level ($x_1(i)$, $1 \leq i \leq n_2$) interact with each other and received without any noise
    \item Those parts of $x_1$ and $x_2$ that are below noise level are truncated and not received at all
\end{itemize}
The key point is how to model the interaction between the bits that are received at the same signal level. In our deterministic model we ignore the carry-over's of the real addition and we model the interaction by the modulo 2 sum of the bits that are arrived at the same signal level. Pictorially the deterministic model for a Gaussian MAC channel is shown in figure \ref{fig:mac} (a). Analogous to the deterministic model for the point-to-point channel, we can write
\beq \mathbf{y}={\bf S^{q-n_1}}\mathbf{x_1} \oplus {\bf S^{q-n_2}}\mathbf{x_2} \eeq where the summation is in $\FF_2$ (modulo 2). Here $\mathbf{x_i}$ ($i=1,2$) and $\mathbf{y}$ are binary vectors of length $q$ denoting transmit and received signals respectively and $\Sbf$ is a $q \times q$ shift matrix.
There is also a relationship between $n_i$'s and the channel gain in dB:
\beq n_i \leftrightarrow \lceil \log  \SNR_i \rceil , \quad i=1,2 \eeq
Note that if one wants to make a connection between the deterministic model and real Gaussian MAC channel (rather than complex) a factor of $\frac{1}{2}$ is necessary.

 Now compared to simple point-to-point case we now have interaction between the digits that receive at the same signal level at the receiver. However, we limit the receiver to only observe the modulo 2 summation of those bits that arrive at the same signal level. In some sense this way of modeling interaction is similar to the collision model. In the collision model if two packets arrive simultaneously at a receiver both are dropped, similarly here if two bits arrive simultaneously at the same signal level the receiver gets only their modulo 2 sum, which means it can not figure out any of them. On the other hand, unlike in the simplistic collision model where the entire packet is lost when there is collision, the most significant bits of the stronger user remain intact. This is reminiscent of the familiar {\em capture} phenomenon in CDMA systems: the strongest user can be heard even when multiple users simultaneously transmit.

%\begin{figure*}%[htp]
%     \centering
%     %%\subfigure[Compact representation]{
%      %% \input{WirelessGraph.pstex_t}
%
%   %%  \hspace{.3in}
%    % \subfigure[Deterministic relay network]{
%       \input{Gaussian_mac_cap.pstex_t}
% \caption{Capacity region of Gaussian MAC channel (solid line). Capacity region of deterministic MAC channel (dashed line)} \label{fig:Gaussian_mac_cap}
%\end{figure*}

% \begin{figure*}%[htp]
%     \centering
%     %%\subfigure[Compact representation]{
%      %% \input{WirelessGraph.pstex_t}
%
%   %%  \hspace{.3in}
%    % \subfigure[Deterministic relay network]{
%       \input{mac2.pstex_t}
% \caption{Pictorial representation of the deterministic model for Gaussian MAC} \label{fig:det_mac}
%\end{figure*}

Now a natural question is how close is the deterministic model to the actual Gaussian model. To answer this question we look at the capacity region of the deterministic MAC. It is easy to verify that the capacity region of the deterministic MAC is the set of non-negative pairs $(R_1,R_2)$ satisfying
\begin{eqnarray}R_2 & \leq & n_2  \\
R_1+R_2 & \leq & n_1 \end{eqnarray}
where $n_i=\log \SNR_i$ for $i=1,2$. This region is plotted with dashed line in figure \ref{fig:mac} (b). In this deterministic model the "carry-over" from one level to the next that would happen with real
addition is ignored. However as we notice still the capacity region is very close to the capacity region of the Gaussian model. In fact it is easy to verify that they are within one bit per user of each other (\emph{i.e.} if a pair $(R_1,R_2)$ is  in the capacity region of the deterministic MAC then there is a pair within one bit per component of $(R_1,R_2)$ that is in the capacity region of the Gaussian MAC). The intuitive explanation for this is that in real addition once two bounded signals are added together the magnitude increases however, it can only become as large as twice the maximum size of individual ones. Therefore the cardinality size of summation is increased by at most one bit. On the other hand in finite-field addition there is no magnitude associated with signals and the summation is still in the same field size as the individual signals. So the gap between Gaussian and deterministic model for two user MAC is intuitively this one bit of cardinality increase.

\subsection{The Deterministic Model for General Networks}
At this point we are ready to explicitly introduce the deterministic model for general wireless relay networks. We model a wireless network $G$ as a set of nodes $V$, where $|V|=N$.

Communication from node $i$ to node $j$ has a non-negative integer
gain\footnote{Some channels may have zero gain.} $n_{(i,j)}$ associated
with it. This number models the channel gain in a corresponding Gaussian setting.
At each time $t$, node $i$ transmits a vector ${\bf x_i}[t] \in
\FF_{2}^q$ and receive a vector ${\bf y_i}[t] \in \FF_{2}^q$ where
$q=\max_{i,j}(n_{(i,j)})$. The received signal at each node is
a deterministic function of the transmitted signals at the other
nodes, with the following input-output relation: if the nodes in the
network transmit ${\bf x_1}[t], {\bf x_2}[t] , \ldots {\bf x_N}[t]$ then the received
signal at node j, $1 \leq j \leq N$ is:
\begin{equation}
\label{eq:channel_model}
{\bf y_{i}}[t]=\sum_{k=1}^N
 {\bf S^{q-n_{k,j}}}{\bf x_{k}}[t]
\end{equation}
for all $1 \leq k \leq N$ and the summation and
multiplication is in $\FF_{2}$.

\begin{figure*}%[htp]
     \centering
     %%\subfigure[Compact representation]{
      %% \input{WirelessGraph.pstex_t}

   %%  \hspace{.3in}
    % \subfigure[Deterministic relay network]{
       \input{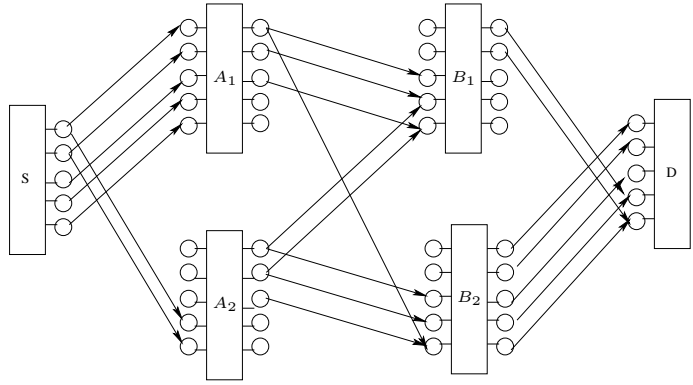}
 \caption{Pictorial representation of Deterministic relay network} \label{fig:RelayModel}

\end{figure*}

The deterministic wireless network can be represented pictorially
and an example is illustrated in Figure \ref{fig:RelayModel}.

\subsection{Related Works}

Finite field addition makes the model much more tractable, and neglecting the 1-bit carryover from one level to the next introduce a small error when the SNR is high. Other works \cite{seven_people_paper} have also exploited the simplicity of finite-field addition over real addition. Aref \cite{Aref_thesis} is one of the earliest works that use
deterministic models for relay networks, and for which he proved a
capacity result for the single-source-single-destination case.
However, his model only captures the broadcast aspect but not the
superposition aspect. This work was later extended to the multicast
setting by Ratnaker and Kramer \cite{RK06}. Aref and El Gamal
\cite{AE82} also computed the capacity of the semi-determinstic
relay channel but only with a single relay. Gupta et al \cite{GB06}
also uses finite-field deterministic addition to model the
superposition property, but they do not have the notion of signal
scale and the channel as sending some of the signal scales to below
noise level. Instead they use random erasures to model noise.
%%Unfortunately it is not possible to connect these models to the Gaussian model.

\section{Single-Source, Single-Destination Network and its Capacity}
\label{sec:MainRes}

Given the deterministic model of Section \ref{sec:DetermisticModel},
we study the information flow for a single source-destination network (unicast).

First we derive the cut-set upper bound on the capacity of this network.

\begin{defn}
A cut, $\Omega$ in the deterministic relay network $G$ with two
distinguished vertices: the source, $S$, and the destination, $D$, is
a split of the vertices into two disjoint sets $\Omega$ and
$\Omega^c$, such that  $S\in \Omega$ and $D\in \Omega^c$.
\end{defn}

For any cut $\Omega$ we define $\Gbf_{\Omega,\Omega^c}$ as the incidence matrix
associated with the bipartite graph with the small nodes of $\Omega$
on the left side and the small nodes of $\Omega^c$ on the right side
and with all edges going from small nodes of $\Omega$ to small nodes
of $\Omega$ based on the equation described in
(\ref{eq:channel_model}). For example in Figure \ref{fig:RelayModel}
consider the cut $\Omega$ that separates $\Omega=\{S,A_1,A_2\}$ from
$\Omega^c=\{B_1,B_2,D \}$ then $\Gbf_{\Omega,\Omega^c}$ is just the incidence matrix of the bipartite graph
between the small nodes on $A_1$ and $A_2$
and  the small nodes on  $B_1$ and $B_2$. Therefore

{\footnotesize
\beq \Gbf_{\Omega,\Omega^c} = \left(
                         \begin{array}{cccccccccc}
                           0 & 0 & 0 & 0 & 0 & 0 & 0 & 0 & 0 & 0 \\
                           0 & 0 & 0 & 0 & 0 & 0 & 0 & 0 & 0 & 0 \\
                           1 & 0 & 0 & 0 & 0 & 0 & 0 & 0 & 0 & 0  \\
                           0 & 1 & 0 & 0 & 0 & 1 & 0 & 0 & 0 & 0 \\
                           0 & 0& 1 & 0 & 0 & 0 & 1 & 0 & 0 & 0  \\
                           0 & 0 & 0 & 0 & 0 & 0 & 0 & 0 & 0 & 0  \\
                           0 & 0 & 0 & 0 & 0 & 0 & 0 & 0 & 0 & 0 \\
                           0 & 0 & 0 & 0 & 0 & 1 & 0 & 0 & 0 & 0  \\
                           0 & 0 & 0 & 0 & 0 & 0 & 1 & 0 & 0 & 0  \\
                           1 & 0 & 0 & 0 & 0 & 0 & 0 & 1 & 0 & 0  \\
                         \end{array}
                       \right)
\eeq
}

Equivalently, $\Gbf_{\Omega,\Omega^c}$ is the transfer matrix from the super vector of all signals transmitted on the nodes in $\Omega$ to the vector of all received signals on the nodes in $\Omega^c$.

Now based on the cut-set bound theorem \cite{CoverThomas91} we have,

%\beq C \leq \max_{p( x_1,\ldots , x_n)} \min_{\Omega}  I \lp \{x_i| i \in
%U_\Omega\};\{y_j| j \in V_\Omega\}|\{x_i| i \in V_\Omega\} \rp  \eeq
%Now in the following lemma we state that the cut values can be
%simultaneously optimized by independent and uniform distribution of
%$x_i$'s in $\FF_{p}^q$ and therefore the min-cut can be expressed
%in terms of the minimum rank of the transfer matrices associated
%with cuts:
\begin{lemma}
\label{lem:MinCut}
The capacity $C$ of any deterministic wireless network $G$ is upper
bounded by \beq C\leq \min_{\Omega}\mathrm{rank}(\Gbf_{\Omega,\Omega^c}) \eeq
\end{lemma}
\begin{proof}
From the cut-set upper bound theorem \cite{CoverThomas91} we have
\beq
C \leq \max_{p( x_1,\ldots , x_n)} \min_{\Omega} I \lp \{x_i| i \in
\Omega\};\{y_j| j \in \Omega^c\}|\{x_i| i \in \Omega^c\} \rp
\eeq
Since the channels are deterministic we can write this as
\beq
C \leq
\max_{p( x_1,\ldots , x_n)} \min_{\Omega} H \lp \{\{y_j| j \in
\Omega^c\}|\{x_i| i \in \Omega^c\} \rp
\eeq
Now note that each of these conditional entropies is at
most equal to the dimension of the range space of the transfer matrix
associated with that cut, achieved when the conditional output is uniformly distributed over its possible values. Now by properties of finite-field arithmetic, this can be simultaneously achieved for {\em all} conditional entropies by choosing independent and uniform
distribution of $x_i$'s in $\FF_{2}^q$. Hence, the cut-set bound can
be expressed in terms of the minimum rank of the transfer matrices
associated with the cuts.
\end{proof}

Now the following main theorem states that the capacity of the wireless deterministic network is equal to its cut-set bound.

\begin{theorem}
\label{thm:Main} If $G$ is a deterministic wireless network
the cooperative capacity of this network from $S$ to $D$ denoted by
$C$ is equal to \beq \label{eq:MainThm}
C=\min_{\Omega}\mathrm{rank}(\Gbf_{\Omega,\Omega^c}) \eeq where the minimum is taken over all cuts
in $G$.
\end{theorem}
\begin{proof}
The proof of this result can be found in the sequel \cite{allerton_paper_part_two} to this paper.
\end{proof}

For the example shown in Figure \ref{fig:RelayModel} the theorem
states that the capacity of this wireless deterministic network is
equal to $5$ which is the value of the cut with
$\Omega=\{S,A_1\}$ and $\Omega^c=\{A_2,B_1,B_2,D\}$. Note
that there are several other tight cuts such as $\Omega=\{S\}$
and $\Omega^c=\{A_1,A_2,B_1,B_2,D\}$.

For wireline networks with unit-capacity edges, the classic max-flow min-cut theorem says that the maximum achievable rate from source to destination is equal to the minimum of the values of the cuts, where the value of a cut is the number of edges crossing it. Theorem \ref{thm:Main} can be viewed as an analogy of this result for our deterministic model, with the cut value being the rank of the transfer matrix. In fact, both the wireline model and our deterministic model are special cases of a more general class of linear deterministic models, where the matrix ${\bf S^{q-n_{k,j}}}$ in equation (\ref{eq:channel_model}) is replaced by a general binary matrix ${\bf G_{k,j}}$. The analysis of this class of model is the focus of \cite{allerton_paper_part_two}, and the main result there is a generalization of both the classic max-flow min-cut theorem and Theorem \ref{thm:Main}.

%Intuitively speaking the maximum flow that can pass through each cut is the same as the number of the linearly independent equations received at the other side of the cut. This motivates us to think of a communication scheme from the source to the destination as forwarding enough number of equations that enables the receiver to decode the transmitted message. Since this type of scheme has a routing nature (forwarding equations) one might suspect that there is a connection between a deterministic relay network and another corresponding wireline network. In fact we show that there is a quite interesting connection between a deterministic relay network and a wireline network called its \emph{wireline backbone} which is described in section \ref{sec:wireline}.

%\section{Connections To Wireline Network}
%\label{sec:wireline}
%%\input{wireline}
%Maybe discuss wireline backbone and its connections here

\section{Connections to Gaussian Relay Networks}
\label{sec:connection_gaussian}
In this section we will discuss some connections between the
deterministic model and the Gaussian model. We will look at two
examples of Gaussian relay networks. In these examples we will show
that a capacity-achieving scheme in the corresponding deterministic
model naturally suggests a scheme in the Gaussian network that
achieves a rate whose gap from its cut-set upper bound is bounded independent of the values of the channel gains. Therefore we have uniformly good approximation of the
capacities of these relay networks, uniform over
all values of the channel gains.

\subsection{Relay channel to within one bit}

\begin{figure*}%[htp]
     \centering
     %%\subfigure[Compact representation]{
      %% \input{WirelessGraph.pstex_t}

   %%  \hspace{.3in}
    % \subfigure[Deterministic relay network]{
       \epsfig{file=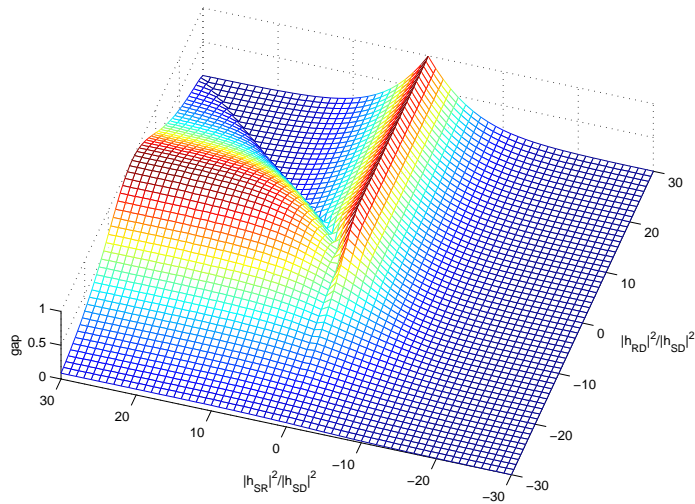,scale=0.5}
 \caption{The gap between cut-set upper bound and achievable rate of decode-forward scheme in Gaussian relay channel} \label{fig:relay}

\end{figure*}

In this section we look at a simple Gaussian network with only one
relay. The capacity of this network has been an open problem for
several decades. Here we will use the deterministic model to find a
near-optimal communication scheme for this network. First we build the
corresponding deterministic model of this relay channel with channel
gains denoted by $n_{SR}$, $n_{SD}$ and $n_{RD}$.  From theorem
\ref{thm:Main} we know that the capacity of this deterministic channel
is equal to {\footnotesize \begin{eqnarray} C_{relay}^d &=&\min \lp
\max(n_{SR},n_{SD}), \max(n_{RD},n_{SD}) \rp \\
\label{eq:det_relay_cap} &=& n_{SD}+\min \lp (n_{SR}-n_{SD})^+,
(n_{RD}-n_{SD})^+ \rp ~ \end{eqnarray} } Note that equation
(\ref{eq:det_relay_cap}) naturally implies a capacity-achieving
scheme for this deterministic relay network: First $n_{SD}$ bits are sent
from the source directly to the destination; then, the remaining $\min
\lp (n_{SR}-n_{SD})^+, (n_{RD}-n_{SD})^+ \rp $ bits can be routed on
the non-interfering signal levels from the source to the relay and
then to the destination. This suggests a decode-and-forward scheme
for the original Gaussian relay channel. If $|h_{SR}|<|h_{SD}|$ then
the relay is ignored and a communication rate equal to
$R=\log(1+|h_{SD}|^2)$ is achievable. If $|h_{SR}|>|h_{SD}|$ the
problem becomes more interesting. In this case we can think of a
decode-forward scheme as described in \cite{cover_elgamal_relay}. Then
by using a block-Markov encoding scheme the following communication
rate is achievable: \beq R = \min \lp \log \lp 1+|h_{SR}|^2 \rp , \log
\lp 1+|h_{SD}|^2+|h_{RD}|^2 \rp \rp \eeq Therefore overall the
following rate is always achievable:
%{\small
\begin{eqnarray} \nonumber  R_{\text{DF}}  = \max \{ \log(1+|h_{SD}|^2)    ,\quad \quad \quad  \quad \quad \quad  \\     \min \lp \log \lp 1+|h_{SR}|^2 \rp , \log \lp 1+|h_{SD}|^2+|h_{RD}|^2 \rp \rp  \} \end{eqnarray}
Now we show that the achievable rate of this communication scheme is
within one bit of the cut-set upper bound of this network for all
channel gains. To do so we should compare this achievable rate by the
cut-set upper bound on the capacity of the Gaussian relay network,
{\small
\begin{eqnarray}
\nonumber C &\leq & \overline{C}= \max_{|\rho| \leq 1} \min \{ \log
\lp 1+(1-\rho^2)(|h_{SD}|^2+|h_{SR}|^2) \rp , \\ && , \log \lp
1+|h_{SD}|^2+|h_{RD}|^2+2\rho |h_{SD}||h_{RD}| \rp \}
\end{eqnarray}
}
Note that if $|h_{SR}|>|h_{SD}|$ then
{\small \beq R_{DF}= \min \lp \log \lp
1+|h_{SR}|^2 \rp , \log \lp 1+|h_{SD}|^2+|h_{RD}|^2 \rp \rp \eeq }and
for all $|\rho|\leq 1$ we have {\small \beq \log \lp
1+(1-\rho^2)(|h_{SD}|^2+|h_{SR}|^2) \rp \leq \log \lp 1+|h_{SR}|^2
\rp+1 \eeq }
 \begin{eqnarray}
\nonumber \log \lp 1+|h_{SD}|^2+|h_{RD}|^2+2\rho |h_{SD}||h_{RD}| \rp
\leq \\ \log \lp 1+|h_{SD}|^2+|h_{RD}|^2 \rp +1
\end{eqnarray}
Hence \beq R_{\text{DF}} \geq \overline{C}_{\text{relay}}-1 \eeq Also
if $|h_{SR}|>|h_{SD}|$, \beq R_{DF}=\log(1+|h_{SD}|^2) \eeq and {\small \beq
\log \lp 1+(1-\rho^2)(|h_{SD}|^2+|h_{SR}|^2) \rp \leq \log \lp
1+|h_{SD}|^2 \rp+1 \eeq } therefore again, \beq R_{\text{DF}} \geq
\overline{C}_{\text{relay}}-1 \eeq

Therefore we showed that the maximum gap between decode-forward achievable rate and the cut-set upper bound on the capacity of Gaussian relay network is at most one bit. However we should point out that even this 1-bit gap is too conservative in many parameter values. In fact the gap would be at the maximum value only if two of the channel gains are exactly the same. Since in a wireless scenario the channel gains differ significantly this happens very rarely. In figure \ref{fig:relay} the gap between the achievable rate of decode-forward scheme and the cut-set upper bound is plotted for different channel gains. In this figure x and y axis are respectively representing the channel gains from relay to destination and source to relay normalized by the gain of the direct link (source to destination) in dB scale. The z axis shows the gap (in bits). There are two main points that one should note in this figure: first the gap is at most one bit which is consistent with what we showed in this section. Second the maximum value happens in some rare cases that two channel gains are exactly equal and on the average the gap is much less than one bit.

\subsection{Diamond network to within two bits}

\begin{figure*}%[htp]
     \centering
       \input{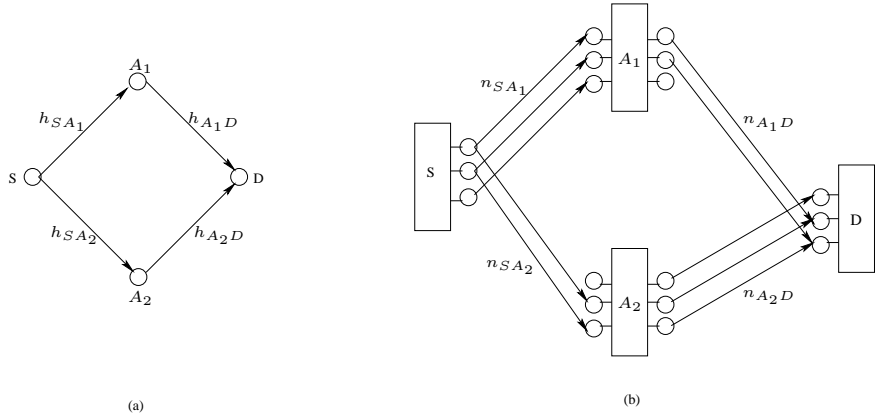}
 \caption{Diamond network with two relays: (a) Gaussian Model, (b) Deterministic Model} \label{fig:diamond}
\end{figure*}

\begin{figure*}%[htp]
     \centering
       \input{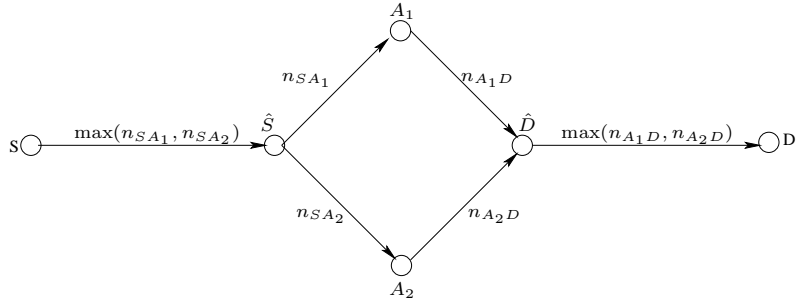}
 \caption{Wireline diamond network. The outgoing links at node $\hat{S}$ are orthogonal; the incoming links at node $\hat{D}$ are orthogonal.} \label{fig:diamond_wired}
\end{figure*}

Consider the diamond Gaussian relay network shown in figure
\ref{fig:diamond}(a). Brett Schein introduced this network in his
thesis \cite{brett_schein} and investigated its capacity. But the
capacity of this network is still an open problem . Here we will
discuss how we can use the deterministic model to approximate the
capacity of this channel within two bits.
First we build the corresponding deterministic model of
this relay channel as shown in figure \ref{fig:diamond}(b). By theorem
\ref{thm:Main} we know that the capacity of this deterministic channel
is equal to

\begin{eqnarray}
\nonumber C_{\text{diamond}}^d &=&\min \{ \max(n_{SA_1},n_{SA_2}),
\max(n_{A_1D},n_{A_2D}) \\ && \quad \quad ,
n_{SA_1}+n_{A_2D},n_{SA_2}+n_{A_1D} \}
\end{eqnarray}
From these constraints it is easy to see that the capacity of the
diamond deterministic network is equal to the capacity of the wireline
network shown in figure \ref{fig:diamond_wired}. By the max-flow
min-cut theorem we know that the capacity of the wireline diamond
network is achieved by a routing solution. It is not difficult to see that the capacity of
the deterministic diamond network can also be achieved mimicking that routing solution by sending
information through non-interfering links from source to relays and
then from relays to destination. A natural analogy of this scheme
(that achieves the capacity of the deterministic diamond network) for
the Gaussian network is the following partial decode-and-forward strategy:
\begin{enumerate}
  \item The source broadcasts two messages, $m_1$ and $m_2$, at rate
  $R_1$ and $R_2$ to relays $A_1$ and $A_2$
  \item Each relay $A_i$ decodes message $m_i$, $i=1,2$
  \item Then $A_1$ and $A_2$ re-encode the messages and transmit them
  via the MAC channel to the destination
\end{enumerate}
Clearly at the end the destination can decode both $m_1$ and $m_2$
with small error probability if, $(R_1,R_2)$ is inside the capacity
region of the BC from source to relays as well as the capacity
region of the MAC from relays to the destination. Assume
$|h_{SA_1}|>|h_{SA_2}|$ then define the following region as the
intersection of BC (from source to relays) and MAC (from relays to
destination):  \begin{eqnarray} \nonumber & \mathcal{R}= \bigcup_{\alpha \in [0,1]}
\{(
R_1, R_2) ~ \text{s.t. }  \quad \quad \\  \label{eq:bc_mac_intersect} & \left\{
                          \begin{array}{ll}
                            0\leq R_1  \leq  \log (1+\alpha|h_{SA_1}|^2),  \\
                            0 \leq R_2  \leq  \log (1+\frac{(1-\alpha)|h_{SA_2}|^2}{\alpha|h_{SA_2}|^2+1}), \\
                            0 \leq R_i  \leq  \log (1+|h_{A_iD}|^2),\quad i=1,2 \\
                            R_1+R_2  \leq  \log \lp 1+ |h_{A_1D}|^2+|h_{A_2D}|^2 \rp
                          \end{array}
                        \right.  \} \end{eqnarray}
Therefore in the Gaussian diamond network
the following communication rate from $S$ to $D$ is achievable: \beq
\label{eq:pdf_rate} R_{\text{PDF}}=\max \{ R_1+R_2|(R_1,R_2) \in
\mathcal{R} \} \eeq Now we will show that the achievable rate of this
partial decode and forward scheme is within two bits of the cut-set
upper bound on the capacity of Gaussian diamond network. To do so
first we define the region $\mathcal{R^*}$ to be
 \begin{eqnarray}
\nonumber & \mathcal{R}^*=\{(R_1^*, R_2^*) ~
\text{s.t. } \quad \quad \quad \\ \label{eq:bc_mac_intersect_upper} &  \left\{
                          \begin{array}{ll}
                            0 \leq R_2^*  \leq \log (1+|h_{SA_2}|^2),  \\
                             R_1^*+ R_2^*  \leq  \log (1+|h_{SA_1}|^2), \\
                            0 \leq R_i  \leq  \log (1+|h_{A_iD}|^2),\quad i=1,2 \\
                            R_1+R_2  \leq  \log \lp 1+ |h_{A_1D}|^2+|h_{A_2D}|^2 \rp
                          \end{array}
                        \right.\}
\end{eqnarray}

Also define,
\beq
\label{eq:pdf_rate_upper} R^*= \max \{ R_1^*+R_2^*|(R_1^*,R_2^*) \in \mathcal{R^*} \}
\eeq
Now we show the following lemma
\begin{lemma}
Consider the rate regions $\mathcal{R}$ and $\mathcal{R}^*$ as
described in (\ref{eq:bc_mac_intersect}) and
(\ref{eq:bc_mac_intersect_upper}). Also assume that
$|h_{SA_1}|>|h_{SA_2}|$ then,
\beq
\mathcal{R} \subseteq \mathcal{R}^*
\eeq
and moreover,
\beq
\label{eq:first_step_diamond} 0 \leq R^*-R_{\text{PDF}} \leq 1
\eeq
where $R_{\text{PDF}}$ and $R^*$ are respectively defined in
(\ref{eq:pdf_rate}) and (\ref{eq:pdf_rate_upper}).
\end{lemma}
\begin{proof}
To show the first part assume $(R_1,R_2) \in \mathcal{R}$. Since this
pair is in the capacity region of BC from source to relays then the
stronger user ($A_1$) should decode both messages and therefore
\beq
R_1+R_2 \leq \log (1+|h_{SA_1}|^2)
\eeq
Now since the last two conditions of $\mathcal{R}$ and $\mathcal{R}^*$
are the same therefore $(R_1,R_2)\in \mathcal{R}^*$ and hence
$\mathcal{R} \subseteq \mathcal{R}^*$.

To prove the second part we show that if $(R_1^*,R_2^*) \in
\mathcal{R}^*$ then $(R_1^*,(R_2^*-1)^+) \in \mathcal{R}$. If $R_2^*
\leq 1$ it is obvious. Otherwise first we find $\alpha^* \in [0,1]$
such that
\beq
\log
(1+\frac{(1-\alpha^*)|h_{SA_2}|^2}{\alpha^*|h_{SA_2}|^2+1}) = R_2^*-1
\eeq
by solving this equation we get
\beq
\alpha^*=\frac{1+|h_{SA_2}|^2-2^{R_2^*-1}}{2^{R_2^*-1}|h_{SA_2}|^2}
\eeq
Now by using the fact that $ |h_{SA_1}| \geq |h_{SA_2}|\geq
2^{R_2^*}-1 $ we have,
\begin{eqnarray*}
\alpha^* & = &
\frac{1+|h_{SA_2}|^2-2^{R_2^*-1}}{2^{R_2^*-1}|h_{SA_2}|^2} \\ &=&
\frac{1+|h_{SA_2}|^2}{2^{R_2^*}|h_{SA_2}|^2}+\frac{1+|h_{SA_2}|^2-2^{R_2^*}}{2^{R_2^*}|h_{SA_2}|^2}
\\ &\geq & \frac{1+|h_{SA_1}|^2}{2^{R_2^*}|h_{SA_1}|^2}
\end{eqnarray*}
therefore we have \beq \log (1+\alpha^* |h_{SA_1}|^2) \geq \log \lp
\frac{1+|h_{SA_1}|^2}{2^{R_2^*}} \rp \eeq Hence,
\begin{eqnarray*}
R_1^* & \leq & \log (1+|h_{SA_1}|^2) - R_2^* \\ &=& \log \lp \frac{1+|h_{SA_1}|^2}{2^{R_2^*}} \rp \\ & \leq & \log (1+\alpha^* |h_{SA_1}|^2)
\end{eqnarray*}
therefore $(R_1^*,(R_2^*-1)^+) \in \mathcal{R}$ and the proof is complete.
\end{proof}
 As the next step we show that $R^*$ is within one bit of the cut-set
 upper bound on the capacity of Gaussian diamond network. First note
 that
\begin{eqnarray}
R^* & \geq & \min \{ \log \lp 1+ \max( |h_{SA_1}|^2,|h_{SA_2}|^2) \rp
\\ & & \quad \quad , \log \lp 1+ |h_{A_1D}|^2+|h_{A_2D}|^2 \rp \\ & &
\quad \quad , \log (1+|h_{SA_1}|^2)+\log (1+|h_{A_2D}|^2) \\ & & \quad
\quad , \log (1+|h_{SA_2}|^2)+\log (1+|h_{A_1D}|^2) \} ~~~~~~
\end{eqnarray}
since the right hand side is achievable. On the other hand the cut-set
upper bound is upper bounded by,
\begin{eqnarray}
\nonumber C_{\text{diamond}} & \leq & \overline{C} \leq \min \{ \log
\lp 1+ |h_{SA_1}|^2+|h_{SA_2}|^2 \rp \\ \nonumber & & , \log \lp 1+
(|h_{A_1D}|+|h_{A_2D}|)^2 \rp \\ \nonumber & & , \log
(1+|h_{SA_1}|^2)+\log (1+|h_{A_2D}|^2) \\ & & , \log
(1+|h_{SA_2}|^2)+\log (1+|h_{A_1D}|^2) \} ~ ~~~~
\end{eqnarray}
Now note that
{\footnotesize
\beq
\log \lp 1+  |h_{SA_1}|^2+|h_{SA_2}|^2 \rp  \leq  \log \lp 1+ \max( |h_{SA_1}|^2,|h_{SA_2}|^2) \rp +1 \eeq
\beq \log \lp 1+ (|h_{A_1D}|+|h_{A_2D}|)^2 \rp   \leq  \log \lp 1+ |h_{A_1D}|^2+|h_{A_2D}|^2 \rp +1 \eeq
}
Therefore
\beq
\label{eq:second_step_diamond}
R^* \geq \overline{C}- 1
\eeq
combining (\ref{eq:first_step_diamond}) and (\ref{eq:second_step_diamond}) we have
\beq
R_{\text{PDF}} \geq \overline{C}- 2
\eeq

Hence the achievable rate of this partial-decode-forward scheme is
within two bits of the cut set upper bound for all values of the channel gains. It
is probably possible to improve this constant gap further by
choosing a more efficient strategy. However, here our goal is to
concretely show it is possible to characterize the high SNR behavior capacity of the relay network by exhibiting a scheme that is within a constant number of bits to capacity no matter how large the channel gains are. Therefore the exact gap  is not fundamentally
important here.

%%\section{Summary and Further Discussions}
%%In this paper we presented a deterministic model for wireless
%%networks. We gave insight on how the proposed model captures the
%%fundamental features of wireless channels by looking at three basic
%%scenarios in wireless networks: point to point, broadcast and multiple
%%access. We also determined the capacity of such a deterministic relay
%%network in general. Finally we looked at a few examples of Gaussian
%%relay networks and we discussed how the intuition obtained from
%%analyzing the corresponding deterministic model can help to find a
%%scheme that achieves within constant number of bits from cut-set upper
%%bound for all channel gains. In general we conjecture that for any
%%Gaussian relay network there is a scheme that achieves the cut-set
%%bound for all channel gains, within a constant number of bits.

\vspace{0.1in} \noindent {\bf Acknowledgements}: The research of D.
Tse and A. Avestimehr are supported by the National Science
Foundation through grant CCR-01-18784. The research of S. Diggavi is
supported in part by the Swiss National Science Foundation NCCR-MICS
center.

\end{document}